\newcommand{\wn}{cm\textsuperscript{-1}}

\newcommand\T{\rule{0pt}{2.6ex}} 
\newcommand\B{\rule[-1.2ex]{0pt}{0pt}} 

\documentclass[review,sort&compress,times]{elsarticle}
\makeatletter
\DeclareRobustCommand*{\bfseries}{%
  \not@math@alphabet\bfseries\mathbf%
  \fontseries\bfdefault\selectfont%
  \boldmath%
}
\makeatother
\usepackage{graphicx}
\usepackage{gensymb}
\usepackage{epstopdf}
\usepackage{multirow}
\usepackage{textcomp}
\usepackage{dcolumn}
\usepackage{hhline}
\usepackage{amsmath}
\usepackage{svn-multi}
\usepackage{bm,upgreek}
\usepackage{color}
\usepackage{hyperref}

\usepackage[modulo]{lineno}
\newcolumntype{d}[1]{D{.}{.}{#1}}
\usepackage{amssymb}

\begin{document}

\begin{frontmatter}

\title{First pressure shift measurement of ozone molecular lines at 9.54 $\upmu$m using a tunable quantum cascade laser}
\author[lerma,PIIM,Fresnel]{Marco~Minissale}
\ead{marco.minissale@univ-amu.fr}
\author[lerma]{Thomas Zanon-Willette}
\author[lerma]{Pascal Jeseck}
\author[lerma]{Corinne Boursier}
\author[lerma]{Christof Janssen}%
\ead{christof.janssen@upmc.fr}
\address[lerma]{LERMA-IPSL, Sorbonne Universit\'{e}, Observatoire de Paris, PSL Research University, CNRS, F-75005 Paris, France}
\address[PIIM]{now at: Aix Marseille Universit\'{e}, CNRS, PIIM UMR 7345, 13397 Marseille, France}%
\address[Fresnel]{now at: Aix-Marseille Universit\'{e}, CNRS, Centrale Marseille, Institut Fresnel UMR 7249, 13013 Marseille, France}%

\begin{abstract}
 Using a free-running distributed-feedback quantum cascade laser (QCL) emitting at 9.54 $\upmu$m, the pressure shift parameters of four intense rovibrational transitions in the $\nu_3$ fundamental band of ozone induced by oxygen (O$_2$), air and the noble gases helium (He), argon (Ar), and xenon (Xe) are obtained by employing second harmonic detection. The experimental analysis comprises a full uncertainty budget and provides line shift data which are traceable to SI. The high density of transitions in the  $\nu_3$  spectral region of ozone make this region particularly difficult to study with more commonly used techniques such as Fourier transform spectroscopy. The comparatively high spectral resolution of the QCL in the MHz range, on the contrary, allows to measure molecular shifts at relatively low pressures (from 2 to 70 hPa), thus reducing the impact of spectral congestion due to pressure broadening of molecular lines. The comparison of our results with published data shows that presently recommended values for the pressure shift are too low in this region. This observation is corroborated by semi-classical calculations using the Robert-Bonamy formalism. A slight negative $J$ dependence, already observed in other ozone vibrational bands, is predicted. Systematic use of our technique could be very useful to support this hypothesis and to make up for the lack of shift parameters for  ozone $\nu_3$ transitions in molecular spectral databases. A subsequent stabilization of the QCL onto an optical frequency comb will open up possibilities to perform metrological measurements of Doppler-free molecular lines.
\end{abstract}
\begin{keyword}

ozone \sep pressure shift \sep noble gases \sep quantum cascade laser

\end{keyword}

\end{frontmatter}

\section{Introduction}

	Ozone (O$_3$) is the tri-atomic allotrope of oxygen and its chemical formula has been identified almost exactly 150 years ago first by Soret and shortly later by Sch{\"o}nbein~\cite{Rubin:2001:The-history_a}. The molecule  plays a pivotal role in Earth's atmosphere. It is a unique absorber of ultraviolet (UV), visible and infrared (IR) radiation, thus greatly influencing the thermal structure of Earth's atmosphere, protecting Earth's surface from energetic solar UV irradiation, and playing a role in global climate - chemistry feedback. Moreover, the molecule is an irritant that is key in atmospheric oxidation chemistry.

Precise determinations of local ozone mixing ratios or abundance profiles in the atmosphere are thus of great interest and can be obtained through spectroscopic measurements in all spectral domains from the UV to the Far-IR. In the light of new studies and more stringent user requirements, the re-evaluation and new measurements of data covering all these domains are under way \cite{OrphalStaehelin:2016:Absorption_a,BarbeMikhailenko:2013:Ozone_a,GuinetMondelain:2010:Laser_a,JanssenBoursier:2016:Line_a,JanssenElandaloussi:2017:A-new-photometric_a,DrouinCrawford:2017:Validation_a}. The spectral region at 10 $\upmu$m is of particular interest for these measurements which is due to the absence of water absorptions in this atmospheric 'window' and the presence of the strongest fundamental ($\nu_3$) of ozone. The inversion of atmospheric spectra, however, requires a precise knowledge of ozone spectral line parameters  such as position, intensity, line broadening, and shifting parameters~\cite{BarbeMikhailenko:2013:Ozone_a,GamacheArie:1998:Pressure-broadening_a,JanssenBoursier:2016:Line_a}, with required relative uncertainties decreasing with the order in which parameters are listed. In most atmospheric applications, line shifting has thus be neglected altogether~\cite{VaranasiChudamani:1990:The-temperature_a}. More recently, however, line shift data have become a concern in atmospheric applications~\cite{Smith:2006:Remote_a}. One reason being the increasing spectral resolution in modern instrumentation that leads to increased requirements, because rather typical uncertainties in shift coefficients of about $10^{-3}\,$cm\textsuperscript{-1}/atm may lead to significant residues in the retrieval of atmospheric spectra~\cite{SchneiderHase:2009:Improving_a}.


The shifting of ozone rovibrational transitions has been studied both experimentally~\cite{SmithRinsland:1988:Measurements_b,SmithDevi:1997:Temperature_a,DeviBenner:1997:Air-Broadening_a,BarbeBouazza:1991:Pressure_a,SokabeHammerich:1992:Photoacoustic_a,SmithRinsland:1994:Measurements_a} and theoretically~\cite[][and references therein]{GamacheArie:1998:Pressure-broadening_a,LavrentievaOsipova:2009:Calculations_a} mostly in the $\nu_1$, $\nu_2$, $\nu_1+\nu_3$, $\nu_1+\nu_2 + \nu_3$, and $3\nu_3$ vibrational bands. For an extensive review of these measurements up to 1998 readers are referred to Ref.~\cite{GamacheArie:1998:Pressure-broadening_a}. Interestingly, line shifts in the $\nu_3$ band of ozone have only been studied non-systematically such that values for just 10 lines are reported in the literature~\cite{SmithRinsland:1988:Measurements_b,SmithDevi:1997:Temperature_a,SokabeHammerich:1992:Photoacoustic_a}. Due to the lack of experimental data, shift parameters for $\nu_3$ lines in the spectroscopic databases are lacking or must be based on some form of extrapolation from data in other bands. At present, all transitions belonging to this band have shift parameters arbitrarily set to $\delta = -7\cdot 10^{-4}$\,\wn / atm in HITRAN \cite{GordonRothman:2017:The-HITRAN2016_a} or to $\delta = 0$ in GEISA~\cite{Jacquinet-HussonArmante:2016:The-2015_a} and in S\&MPO~\cite{BabikovMikhailenko:2014:SMPO--An_a}. Nevertheless, shifting parameters are vibration dependent~\cite{RobertBonamy:1979:Short_a}, show a large variability~\cite{SmithRinsland:1988:Measurements_b,SmithDevi:1997:Temperature_a,DeviBenner:1997:Air-Broadening_a,BarbeBouazza:1991:Pressure_a,SokabeHammerich:1992:Photoacoustic_a,SmithRinsland:1994:Measurements_a}, and the validity of these attributions thus requires experimental verification.

%
%
%


In this paper, we present experimental data of line shifts of O$_3$ in the $\nu_3$ region induced by the different gases O$_2$, air, He, Ar, and Xe. The results are obtained using QCL technology, instead of the more commonly used Fourier Transform Spectroscopy (FTS) for high-resolution spectroscopic measurements.
Compared to previous line shift studies
\cite{SmithRinsland:1988:Measurements_b,SmithDevi:1997:Temperature_a,DeviBenner:1997:Air-Broadening_a,BarbeBouazza:1991:Pressure_a,SokabeHammerich:1992:Photoacoustic_a,SmithRinsland:1994:Measurements_a} who mostly report the observed scatter as an estimate of the associated measurement uncertainty, we here provide a full uncertainty analysis with metrological traceability.

We recall that SI units for line shifts are Hz/Pa, when line positions are measured in units of frequency and we use the derived unit of kHz/Pa throughout the paper along with values in \wn/atm and MHz/Torr, which are the more traditional units in the radiative transfer or physical chemistry communities. The different units are linked to each other by
\begin{equation}\label{eq:units1}
	1\, \dfrac{\mathrm{kHz}}{\mathrm{Pa}}=\dfrac{72375}{21413747}\, \dfrac{\mathrm{\wn}}{\mathrm{atm}}\simeq 3.380 \cdot 10^{-3}\, \dfrac{\mathrm{\wn}}{\mathrm{atm}}
\end{equation}
and
\begin{equation}\label{eq:units2}
	1\, \dfrac{\mathrm{kHz}}{\mathrm{Pa}}=\dfrac{4053}{30400}\, \dfrac{\mathrm{MHz}}{\mathrm{Torr}}\simeq 0.1333\,\dfrac{\mathrm{MHz}}{\mathrm{Torr}} .
\end{equation}

The paper is organized as follows. We first present the underlying theory for the line shift calculations, before the experimental set-up and the method are described in the following section. Then, we present and discuss the pressure shift measurements. In the conclusion and outlook section we finally summarize the results and propose further spectroscopic applications.

\section{Line shift calculation}\label{sec:theory}

Line shifts have been calculated using the semi-classical formalism of Robert and Bonamy (RB, hereafter) \cite{RobertBonamy:1979:Short_a}. The method, as well as the advantages of using the (RB) formalism rather than the older Anderson-Tsao-Curnutte (ATC) theory \cite{Anderson:1949:Pressure_a,TsaoCurnutte:1962:Line-widths_a}  is well described by Lynch \emph{et al.} \cite{LynchGamache:1996:Fully_a}.  Hence, we present only a short account for providing a basic understanding and explaining the approximations made in the calculations.

The line shift $\delta_{if}$ for the transition $f \leftarrow i$ is calculated using the following equation:
\begin{equation}\label{eq:shiftRB}
		\delta_{if}=\frac{n_2 \bar{v}}{2 \pi c}\left<  \int_{r_0}^\infty 2 \pi r_c \,\mathrm{d}r_c \left(\frac{v_c^\prime}{\bar{v}}\right)^2 \sin S_{\!1} e^{-\Re (S_2)} \right>_2
\end{equation}
where $n_2$ is the perturber number density, $\bar{v}$  the mean relative velocity of the two colliders, $c$ the speed of light, and following  the framework of the RB formalism \cite{RobertBonamy:1979:Short_a}, the parabolic trajectory is described by the distance of closest approach $r_c$ and the effective velocity at this distance $v_c^\prime$, $ r_0 $ being the minimum value of $r_c$ obtained for  a zero impact parameter.  $ \left< \dots \right>_2$ means an average over the rotational states of the perturber, $\Re$ means real  part, and $S_{\!1}$ and $S_{\!2}$ are the usual scattering terms depending on the intermolecular potential including electrostatic and atom-atom terms.

To obtain this equation, the following assumptions have been made:
\begin{itemize}
\item	following the original paper of RB~\cite[][see Eq.~(13)]{RobertBonamy:1979:Short_a}, the  imaginary part $ \Im (S_2)$ has been neglected. Lynch \emph{et al.}~\cite{LynchGamache:1996:Fully_a} have shown that this can lead to an error of about 10\,\% in the case of N$_2$ broadening of water lines.
\item	The $S_{\!1}$ term that depends only on the isotropic part of the interaction
potential is given by :
\begin{equation}\label{eq:S1}
		S_1=-\frac{\alpha_2 \bar{v}}{r_c^5v_c^\prime}\frac{3 \pi}{8 \hbar}\left[\frac{3}{2}\frac{I_1 I_2}{I_1+I_2}\left(\alpha_{1f}-\alpha_{1i}\right) + \left(\mu_f^2 -\mu_i^2\right)\right] ,
\end{equation}
where $\alpha_1 $ and $\alpha_2$ are ozone and perturber polarizabilities, $I_1$ and $I_2$ the respective first ionization energies and $\mu_f$ and $\mu_i$ the ozone electric dipole moment in the inital ($\nu_3=0$) and final $(\nu_3=1)$ vibrational states.
\item	$\Re (S_2)$ is calculated following previous studies of line broadening in ozone \cite{BouazzaBarbe:1993:Measurements_a}, taking into account an  electrostatic interaction potentiel up to the quadrupole-quadrupole term and a Lennard-Jones type atom-atom potential:
\begin{align}\label{eq:atom-atom}
		V_{at-at}&= 4  \sum_{i,j}\epsilon_{ij}\left[\left(\frac{\sigma_{ij} }{r_{ij}}\right)^{12}-\left(\frac{\sigma_{ij} }{r_{ij}}\right)^6\right]\nonumber \\
		&= \sum_{i,j}\left( \frac{d_{ij} }{r_{ij}^{12}}-\frac{e_{ij} }{r_{ij}^6}\right)
\end{align}
where  $\epsilon_{ij}$ and $\sigma_{ij}$ are the Lennard-Jones parameters for the interaction of the $i^{th}$ atom of the active molecule with the $j^{th}$ atom of the perturber.
\end{itemize}
The $S_1$ term has been calculated using the ionization potentials listed in the CRC handbook of physics and chemistry \cite{Haynes:2010:CRC-Handbook_a}. Rare gas polarizabilities and those of N$_2$ and O$_2$ are also given in the CRC handbook \cite{Haynes:2010:CRC-Handbook_a}. The vibrationally dependent dipole moment and polarizability of ozone have respectively been taken from Refs.~\cite{FlaudCamy-Peyret:1990:Atlas_a} and \cite{LavrentievaOsipova:2009:Calculations_a}. 
 For convenience, we summarize the values of the molecular parameters used in the calculations in Table~\ref{tab:molpar}.
\begin{table*}[th!]
\caption{Molecular parameters for the calculation of the pressure shift. Values are given in non-SI units ($1 \AA = 10^{-10}$\,m, $ 1 \mathrm{D} = 8.478353552(52) \cdot 10^{-30}$\, C m).\label{tab:molpar}}
\footnotesize
\begin{center}
\begin{tabular}{lllll}\hline
\T \B  Molecule or atom	& Ionisation energy$^a$	$I$ (eV)	 &	Polarizability$^b$ $\alpha$ (\AA$^3$)	&Dipole moment$^c$ $\mu$ (D)	&	Quadrupole moment$^d$ $Q$ (D \AA )\\ 	\hline
\T He		&	24.5874	 &		0.2051         	 &              		& 				\\
   Ar		&	15.7596	 &		1.6411         	 &              		&				\\
   Xe		&	12.1298	 &		4.044          	 &              		&				\\
   N$_2$		&	15.5808	 &		1.7403 		 	 &					&	$-1.3$		\\
   O$_2$		&	12.0697	 &		1.5689 			 &					&	$-0.39$		\\
   O$_3$		&	12.43	 &	$\alpha_f-\alpha_i=	0.04$ &$\mu_i = 0.5333$ 		&	$Q_{aa}=-1.4$	\\
 			&			 &	 					 &$\mu_f=0.5309$		&	$Q_{bb}=-0.7$	\\
 \B			&			 &						 &					&	$Q_{cc}=2.1$	\\ 	\hline
 \multicolumn{5}{l}{ \T $^a$ Ref.~\cite{Haynes:2010:CRC-Handbook_a} }\\
 \multicolumn{5}{l}{$^b$ Ref.~\cite{Haynes:2010:CRC-Handbook_a}, except for O$_3$ where value for $\alpha_f-\alpha_i$ is taken from Ref.~\cite{LavrentievaOsipova:2009:Calculations_a} }\\
 \multicolumn{5}{l}{$^c$ Ref.~\cite{FlaudCamy-Peyret:1990:Atlas_a} }\\
 \multicolumn{5}{l}{$^d$ For non-linear molecules tensorial components around principal axes are given}\\
%
\hline
\end{tabular}
\end{center}
\end{table*}
Atom-atom interaction parameters are listed separately in Table~\ref{tab:atompar}. Based on Eq.~(\ref{eq:atom-atom}),
values of $d_{\mathrm{OO}}$ and $e_{\mathrm{OO}}$ for the ozone molecule have been calculated from the corresponding $\sigma_{\mathrm{OO}}$ and $\epsilon_{\mathrm{OO}}$ values that have been extracted from the parameters given in Ref.~\cite{BouazzaBarbe:1993:Measurements_a} for O$_3-$O$_2$ and O$_3-$N$_2$ pairs, using standard combination rules. They have then been combined with values for N$_2$ \cite{Bouanich:1992:Site-site_a}, O$_2$ \cite{Bouanich:1992:Site-site_a}, Ar \cite{SalemBouanich:2005:Helium-_a}, He \cite{SalemBouanich:2005:Helium-_a} and Xe \cite{PolingPrausnitz:2000:The-Properties_a}. Molecular trajectory calculations were made using an isotropic 6-12 Lennard Jones potential whose $\epsilon$ and $\sigma$ parameters are also given in Table~\ref{tab:atompar}.

\begin{table}[th!]
\caption{Atom-atom interaction parameters as well as LJ parameters for the trajectory calculation.\label{tab:atompar} }
\footnotesize
\begin{center}
\begin{tabular}{lll}\hline
\T Molecule -	&  LJ parameters for atom-atom pair ($e_{ij},\ d_{ij}$)&	\\
\T perturber pair	& and for isotropic potential ($\sigma,\ \epsilon$) &	Reference\\ 	\hline

\T 		O$_3$-N$_2$ 		& $d($O-N$) = 3.65\cdot 10^{-15}$\,J $\AA^{12}$ 	& O$_3$: \cite{BouazzaBarbe:1993:Measurements_a} \\
						& $e($O-N$ )= 3.10\cdot 10^{-18}$\,J $\AA^{6}$ 	& N$_2$: \cite{Bouanich:1992:Site-site_a} \\
						& $\epsilon/k_{\mathrm{B}} = 114$\,K 						& O$_3$: \cite{Warnatz:1978:Calculation_a} \\
						& $\sigma   = 3.95$\,\AA 					& N$_2$: \cite{PolingPrausnitz:2000:The-Properties_a} \\
 		O$_3$-O$_2$ 		& $d($O-O$) = 2.44\cdot 10^{-15}$\,J $\AA^{12}$	& O$_3$: \cite{BouazzaBarbe:1993:Measurements_a} \\
						& $e($O-O$) = 2.75\cdot 10^{-18}$\,J $\AA^{6}$ 	& O$_2$: \cite{Bouanich:1992:Site-site_a} \\
						& $\epsilon/k_{\mathrm{B}} = 139$\,K 						& O$_3$: \cite{Warnatz:1978:Calculation_a} \\
						& $\sigma   = 3.79$\,\AA 					& O$_2$: \cite{PolingPrausnitz:2000:The-Properties_a} \\
		O$_3$-Ar 			& $d($O-Ar$) = 9.34\cdot 10^{-15}$\,J $\AA^{12}$	& O$_3$: \cite{BouazzaBarbe:1993:Measurements_a} \\
						& $e($O-Ar$) = 6.61\cdot 10^{-18}$\,J $\AA^{6}$ 	& Ar: \cite{SalemBouanich:2005:Helium-_a} \\
						& $\epsilon/k_{\mathrm{B}} = 145$\,K 						& O$_3$: \cite{Warnatz:1978:Calculation_a} \\
						& $\sigma   = 3.81$\,\AA 					& Ar: \cite{SalemBouanich:2005:Helium-_a} \\						
		O$_3$-He 			& $d($O-He$) = 5.45\cdot 10^{-16}$\,J $\AA^{12}$	& O$_3$: \cite{BouazzaBarbe:1993:Measurements_a} \\
						& $e($O-He$) = 8.80\cdot 10^{-19}$\,J $\AA^{6}$ 	& He: \cite{SalemBouanich:2005:Helium-_a} \\
						& $\epsilon/k_{\mathrm{B}} = 44$\,K 						& O$_3$: \cite{Warnatz:1978:Calculation_a} \\
						& $\sigma   = 3.38$\,\AA 					& He: \cite{SalemBouanich:2005:Helium-_a} \\						
		O$_3$-Xe 			& $d($O-Xe$) = 3.43\cdot 10^{-14}$\,J $\AA^{12}$	& O$_3$: \cite{BouazzaBarbe:1993:Measurements_a} \\
						& $e($O-Xe$) = 1.50\cdot 10^{-17}$\,J $\AA^{6}$ 	& Xe: \cite{PolingPrausnitz:2000:The-Properties_a} \\
						& $\epsilon/k_{\mathrm{B}} = 204$\,K 						& O$_3$: \cite{Warnatz:1978:Calculation_a} \\
						& $\sigma   = 4.08$\,\AA 					& Xe: \cite{PolingPrausnitz:2000:The-Properties_a} \\						
\hline
\end{tabular}
\end{center}
\end{table}
\section{Experimental setup}\label{sec:experimental}

\begin{figure}[t]
\centering
\includegraphics[width=7.6cm]{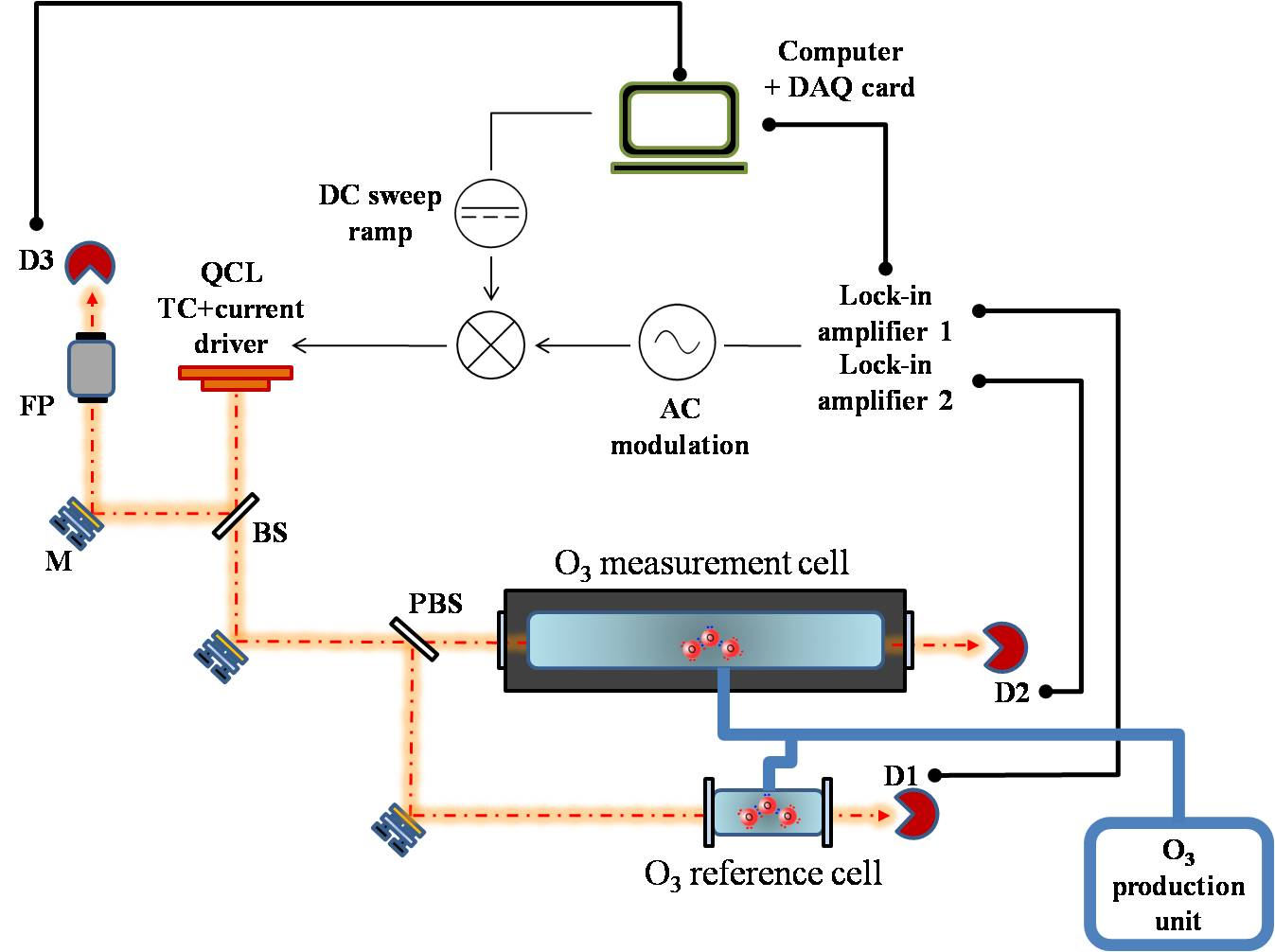}
\caption{Schematic of the experimental setup: QCL, distributed feedback quantum cascade laser; BS, beam splitter; PBS, polarizing beam splitter; FP, Fabry-P\'{e}rot \'{e}talon; D, detector; DAQ card, data acquisition card; TC, temperature controller.} \label{fig:fig1}
\end{figure}

\figurename~\ref{fig:fig1} shows a scheme of the experimental setup, emphasizing on the optical components. Our electromagnetic source is a distributed feedback quantum cascade laser (DFB-QCL,  \emph{Alpes Laser}), centered at 1049 cm$^{-1}$ (9.54\,$\upmu$m) and covering the spectral between 1046.2 and 1053.1\,\wn . The QCL operates at temperatures ranging from $-25$ to $+20\,^{\circ}$C at a maximum current of 1.48\,A. The threshold current changes from 0.88\,A at $-25\,^{\circ}$C to 1.31\,A at $+20\,^{\circ}$C. The QCL is driven by a home-made, low-noise stabilized current source while temperature is stabilized to better than $\pm$0.1\,K using a Peltier module controlled by a thermoelectric cooler (TC-3, \emph{Alpes Laser}). Heat generated by the Peltier element is dumped into a closed-circuit water chiller (ThermoCube, \emph{Solid State cooling}). Moreover, in order to minimize the thermal resistance and to allow efficient dissipation of heat, the laser housing is fixed on a monolithic copper support which is screwed on the optical table. Under typical operation conditions of $T=-10\,^{\circ}$C and $I=1.3\,$A, the QCL emits a power of $\sim$ 25\,mW. It has a tuning rate of $d\nu/dI \sim 5.6 \cdot 10^{-3}$ cm$^{-1}$/mA or 170 MHz/mA.

The output of the QCL can be split by a removable 50-50$\%$ ZnSe beamsplitter (BS) and sent to the Fabry-P\'{e}rot etalon (used for calibration measurements) and, through a fixed BS, to two absorption cells.
The cylindrical absorption cell, which serves as a reference cell (see \figurename~\ref{fig:fig1}), is made of glass onto which two wedged 
BaF$_2$ windows are glued. Attached to the cell is a glass finger (not shown in \figurename~\ref{fig:fig1}) that can be used as cold trap for the storage of liquid ozone. The cell has a base length of 15\,cm and a diameter of 25 mm.

The second absorption cell (denoted "measurement cell" in \figurename~\ref{fig:fig1}) is a cylindric Teflon coated stainless steel cell mounted on two ceramics feet.  The roughly 40 cm long and 50\,mm diameter cell is positioned in an insulating aluminum box. Both, the cell and the box, are equipped with wedged (angle 0.5$^{\circ}$) BaF$_2$ windows on either side. Each BaF$_2$ window decreases the beam power by about 6\,\%. The presence of a slight wedge allows to minimize accidental etalon effects. 
The cell temperature is measured through seven Pt100 RTDs (platinum resistance temperature detectors) integrated into the metallic cell body at $\sim$ 1\,cm from the inner wall surface. 

Both cells are connected to the ozone production system (O$_3$-PS) via two glass-metal transitions. Similar ozone systems have been described elsewhere \citep{Ja06, Ja11} and we give only a very short account here. Briefly, the all-glass system is made of Pyrex, mostly tubing with 8\,mm inner diameter. It connects through three glass-metal transitions to the oil-free turbo molecular pump (TSU 180 H, \emph{Pfeiffer}) that is backed up by an oil-free membrane pump (MD-4, \emph{Vacuubrand}), to the oxygen supply, and to two capacitive pressure transducers. The lower pressure range is covered by a temperature stabilized 10 Torr head (\emph{MKS} 690A), the higher range by a non-stabilized 1000 hPa sensor (\emph{MKS} 122A). Valves are all glass with Teflon fittings (\emph{Glass Expansion}).

Ozone is generated from oxygen with purity of 99.9995\,\% (Alphagaz~2, \emph{Air Liquide}) in the discharge reactor, by adhering to the following procedure: First, the oxygen gas is added to the reactor volume. Then, the lower part of the reactor including the electrodes are immersed into LN$_2$ and a RF discharge is maintained until the pressure has decreased by about 50\,\% due to ozone condensing on the reactor walls. Finally, the remaining oxygen is pumped away, before the LN$_2$ bath is removed to release the condensed ozone into the gas phase.
The base pressure in the two cells is $\sim 1$\,mPa and is controlled through the temperature stabilized pressure transducer.

The output beams from the two cells are focused onto LN$_2$-cooled HgCdTe detectors (\emph{Infrared Associates}, \emph{AeroLaser}) making use of two ZnSe lenses. The photo-currents of the two detectors are pre-amplified by transimpedance amplifiers, then measured with two lock-in amplifiers (SR830, \textit{Stanford Research Systems}).

\begin{figure}[t]
\centering
\includegraphics[width=8.6cm]{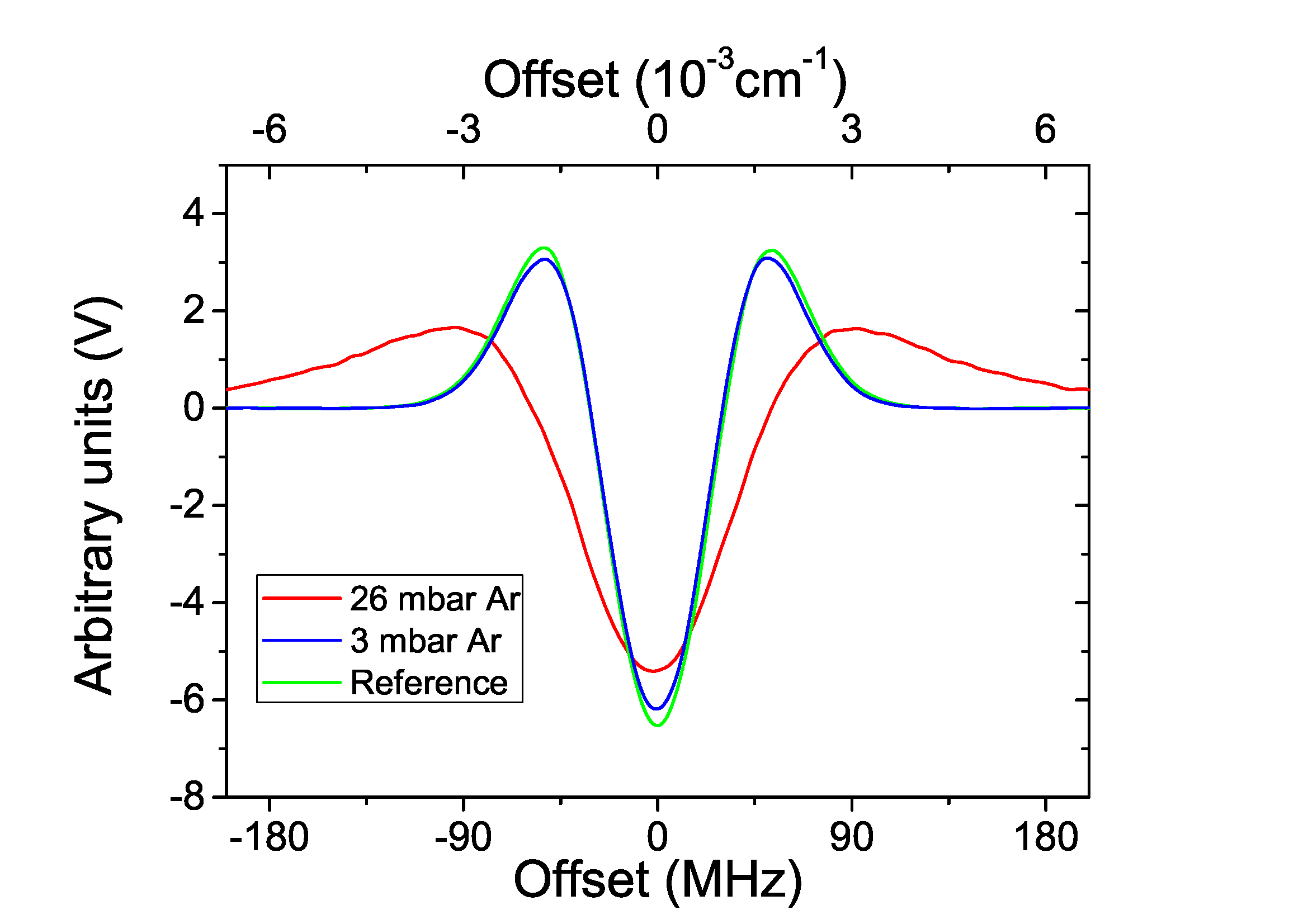}
\caption{Second harmonic spectra showing the pressure shift of an ozone line at three different pressures of Ar (0, 3 and 26 mbar).} \label{fig:fig2a}
\end{figure}

The measurements have been performed using the frequency modulation technique and recording the second derivatives of the signals, as shown in \figurename~\ref{fig:fig2a} for the case of O$_3$-Ar measurements. Shifts are obtained as differences in the minima of the second derivatives. The frequency modulation and the necessity to sweep over a large spectral region containing the target lines, requires to add the sum of a rapid sine-wave modulation and a slow linear voltage ramp to the laser's base current.
The linear ramp is created by a multi purpose data acquisition card (DAQ PCI-6281, \emph{National Instruments}) onboard a PC. The sine-wave modulating signal (at 8.45\,kHz) is generated by one of the lock-in amplifiers. Output signals of the lock-in amplifiers are then recorded by the same DAQ card.

For defining the frequency scale, we use ozone absorption lines together with the signal of a Fabry-P\'{e}rot interferometer (FP in \figurename~\ref{fig:fig1}) with a free spectral range of 240\,MHz ($8.0 \cdot 10^{-3}$\,\wn ).

\subsection{Experimental methods}\label{sec:methods}

For the shift measurements, we fill the reference cell with $\sim (47 \pm 20)$\,Pa of pure ozone. 
In a trap of the O$_3$-PS, we mix a fixed amount of ozone with the perturber gas (PG, hereafter). The measurement cell is filled with this mixture through an adiabatic expansion. Once both cells are loaded, second derivative spectra are acquired continuously and the time difference between second derivative peaks is calculated using a homemade \textit{Labview} routine.

A pair of spectra centered at the target line was acquired at a rate of 1 Hz, due to a triangular sweep at the same frequency. Between 200 and 600 individual spectra were used to determine the average pressure induced time shift in the peak position, corresponding to acquisition times between 100 and 300\,s for one data point.
The time difference is then converted into a frequency shift using separate calibration measurements based on FP etalon spectra.

Once the measurement has been performed for a given initial pressure, we partially empty the measurement cell and repeat the measurements at a lower pressure. This method, which avoids waste, has been employed with the more expensive perturber gases, such as He and Xe. With Ar, O$_2$ and air, the cell is completely evacuated and refilled after a run, such that the amount of ozone remains constant, and only the quantity of PG is changed. Evidently, the second method has the advantage of keeping the ozone amount constant from one measurement run to the next.

We measure a shift
\begin{equation}\label{eq:p-shift}
\Delta=\Delta^{m}_{s}-\Delta^{r}_{s}+\Delta_g,
\end{equation}
where $\Delta^{m}_{s}$ and $\Delta^{r}_{s}$ are ozone self shifts in the measurement and the reference cell, respectively, and where $\Delta_g$ is the shift due to the perturber gas.
By use of the corresponding pressure shift coefficients $\delta $,
Eq.~(\ref{eq:p-shift}) takes the following form
\begin{equation}\label{eq:p-shift2}
\Delta = \delta_{s} \left(-p^{r}_{\mathrm{O_3}} + p^{m}_{\mathrm{O_3}}\right) + \delta_g p_g ,
\end{equation}
where the $\delta_s $ and $\delta_g$ are the self and foreign gas shift coefficients, respectively, and $p_g$, $p^r_{\mathrm{O_3}}$, and $p^m_{\mathrm{O_3}}$  the respective partial pressures of the perturber gas and of ozone in the reference and measurement cells.
If we express the partial pressure $p_g$ of the perturber gas in terms of the measured quantities, the total pressure $p$ and the partial pressure of ozone in the measurement cell $p^{m}_{\mathrm{O_3}}$, Eq.~(\ref{eq:p-shift2}) can be arranged to give:
\begin{equation}\label{eq:p-shift3}
\Delta^{\prime} =  \Delta + \delta_{s} \left( p^{r}_{\mathrm{O_3}} -  p^{m}_{\mathrm{O_3}}  \right) = \delta_g (p -   p^{m}_{\mathrm{O_3}} ) = \delta_g p^{\prime} ,
\end{equation}
where $\Delta^{\prime}$ is the observed frequency shift corrected for self-shift effects and $p^{\prime}$ the PG partial pressure in the measurement cell. When we utilize the second acquisition method and keep $p^{r}_{\mathrm{O_3}} \simeq  p^{m}_{\mathrm{O_3}}$, self shift effects can be neglected.
In this case, a plot of shifts versus $p^{\prime}$ yields a straight line through the origin, with the perturber shift coefficient $ \delta_g $ as the slope.

The ozone partial pressure being varied proportional to the total pressure in the measurement cell, the effect of partial ozone pressures does not cancel when using the first acquisition method, however. Since we keep the ozone mixing ratio $\chi = p^{m}_{\mathrm{O_3}} / (p^{m}_{\mathrm{O_3}} + p_g) $ constant during these measurements, the analysis of the $\Delta$ versus $p$ data yields a straight line
\begin{equation}\label{eq:p-shift_mixingratio}
\Delta = -\delta_{s} p^{r}_{\mathrm{O_3}} + \left(1 -  \chi \left(1- \frac{\delta_s}{ \delta_g} \right) \right) \delta_g  p,
\end{equation}
which has a slope modified by the ratio of self to foreign gas shift parameters $\delta_s / \delta_g$ and the ozone mixing ratio $\chi$. In principle, the offset of the line  $-\delta_{s} p^{r}_{\mathrm{O_3}} $ allows for determining the ozone self shift.

Perturber gases (PG) were used without further purification and had manufacturer (\emph{Air Liquide}) certified impurities of less than  $ 5$\,$\upmu$mol/mol (He, Xe, O$_2$, and synthetic air) or  $ 1$\,$\upmu$mol/mol (Ar). The bottle of synthetic dry air was composed out of 79\,\% of N$_2$ and 21\,\% of O$_2$.

\section{Results}\label{sec:results}

\begin{figure}[t]
\centering
\includegraphics[width=8cm]{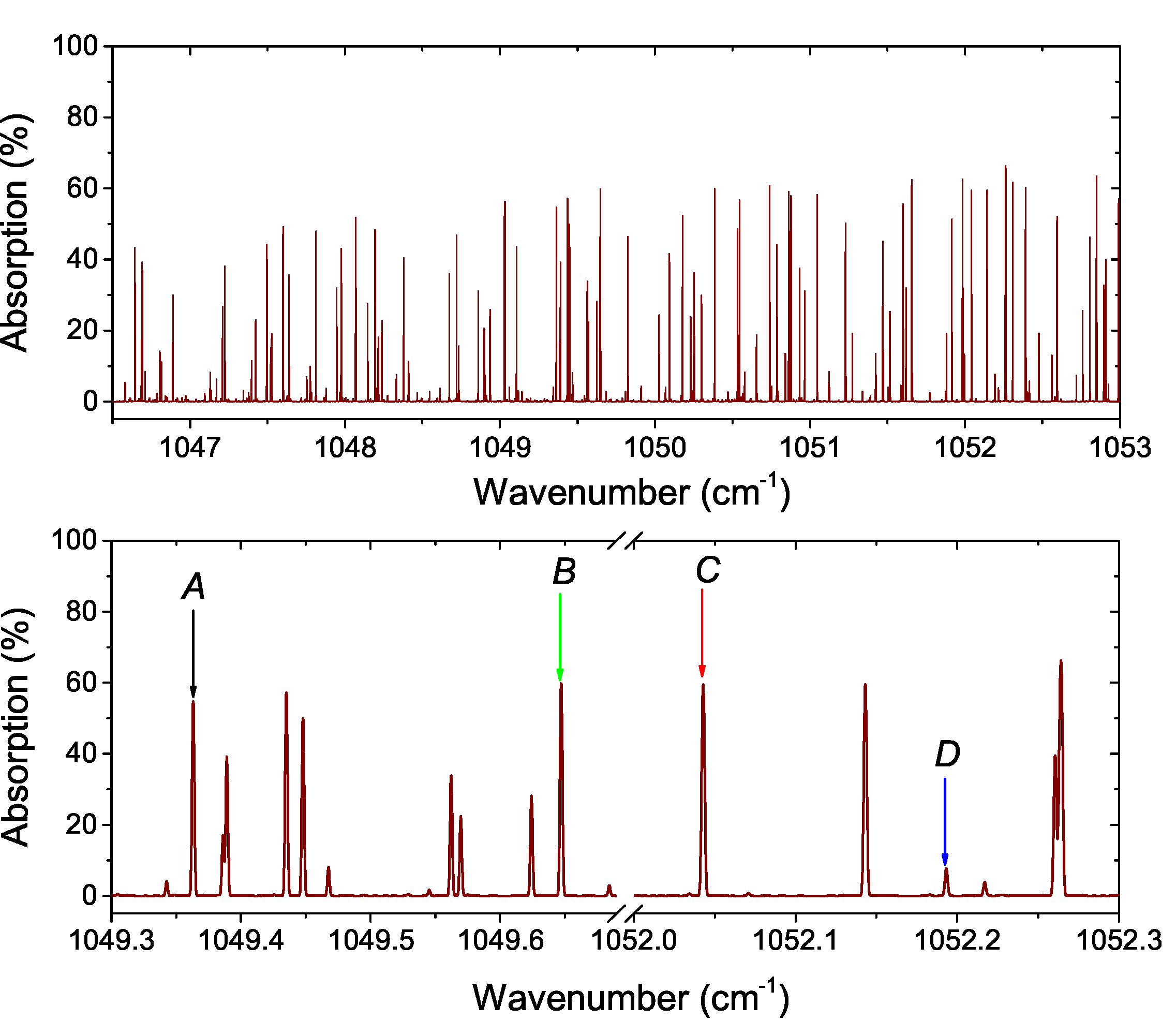}
\caption{Upper panel: ozone absorption in the tuning range of our QCL. The absorption signal is calculated using the HITRAN2016 data. Lower panel: Measured absorption spectrum giving a zoom on the two spectral windows ($1049.3-1049.6$\,\wn and $1052-1052.3$\,\wn), where the four selected lines are situated.\label{fig:fig3}}
\end{figure}

\begin{table*}[th!]
\caption{Pressure shift parameter measurements for lines in the $Q$ and $R$-branches of the $\nu_3$ band as a function of perturber gas (O$_2$, air, Xe, Ar, and He). Values are given in kHz/Pa with standard uncertainties in parentheses.\label{tab:tab1}}
\footnotesize
\begin{center}
\begin{tabular}{lcrrrcrrrcd{5}d{5}d{5}d{5}d{5}d{5}}
\hline
    \multicolumn{10}{l}{\T Transition$^a$}                                                                                          & \multicolumn{6}{l}{\B}      \\\cline{1-9}
\T    &Line position  &  \multicolumn{3}{l}{Upper state}   &&   \multicolumn{3}{l}{Lower state}                                && \multicolumn{6}{l}{Pressure shift parameter $\delta$ (kHz / Pa)} \B \\\cline{3-5}\cline{7-9}\cline{11-16}
\T   &  \multicolumn{1}{l}{ (\wn) }   & $J^\prime$ & $K_a^\prime$ & $K_c^\prime$ && $J^{\prime\prime}$ & $K_a^{\prime\prime}$ & $K_c^{\prime\prime}$ && \multicolumn{1}{l}{He} & \multicolumn{1}{l}{Ar} & \multicolumn{1}{l}{Xe} & \multicolumn{1}{l}{ O$_2$} &  \multicolumn{1}{l}{N$_2$} & \multicolumn{1}{l}{air} \B \\
	\hline																																																					          				 
\T &1037.43368 & 16 &  8 & 9  && 16 &  8 &  8 &&      						  &     						 & 							     &                               &  0.50\textsuperscript{\emph{b}} &         						        \\
$A$&1049.36312 &  9 &  1 & 8  &&  8 &  1 &  7 &&       					   & -0.54(03)\textsuperscript{*} & -0.85(04) \textsuperscript{*}  & -0.54(03)\textsuperscript{*}  & -0.49(06)\textsuperscript{*}        &  -0.50(05)\textsuperscript{*}          \\
$B$&1049.62425 & 13 &  7 & 6  && 12 &  7 &  5 &&       					      & -0.64(05)\textsuperscript{*} & -0.83(05) \textsuperscript{*}  & -0.59(04)\textsuperscript{*}  & -0.71(08)\textsuperscript{*}        &  -0.69(06)\textsuperscript{*}          \\
$C$&1052.04284 & 13 &  2 & 11 && 12 &  2 & 10 && -0.11(04)\textsuperscript{*} & -0.62(03)\textsuperscript{*} & -0.83(03) \textsuperscript{*}  & -0.60(03)\textsuperscript{*}  & -0.64(07)\textsuperscript{*}        &  -0.63(05)\textsuperscript{*}          \\
$D$&1052.19324 & 24 & 11 & 14 && 23 & 11 & 13 && -0.13(05)\textsuperscript{*} & -0.72(05)\textsuperscript{*} & -0.98(04) \textsuperscript{*}  & -0.58(04)\textsuperscript{*}  & -0.57(05)\textsuperscript{*}        &  -0.57(04)\textsuperscript{*}          \\
\\
   &            &    &    &    &&    &    &   &&        					  &    	 					     & 							      &                               & -2.40\textsuperscript{\emph{b}}     &       								      \\
   &1071.08345 & 47 &  4 & 43 && 46 &  4 & 42 &&      					    &     						   &								&                		       & -0.95(09)\textsuperscript{\emph{c}} &  -1.01(03)\textsuperscript{\emph{c}}   \\
   &1071.12318 & 51 &  3 & 48 && 50 &  3 & 47 &&      					    &     						   &								&                			   & -0.71(18)\textsuperscript{\emph{c}} &  -0.61(15)\textsuperscript{\emph{c}}   \\
   &1071.91683 & 49 &  4 & 45 && 48 &  4 & 44 &&      					    &      						   &							    &                               & -0.83(09)\textsuperscript{\emph{c}} &  -0.92(03)\textsuperscript{\emph{c}}   \\
   &            &    &    &    &&    &    &   &&      						  &      						 & 							      &                               & 							        &  -0.68(03)\textsuperscript{\emph{d}}   \\
   &1072.64776 & 51 &  4 & 47 && 50 &  4 & 46 &&      					    &     						   &								&                	              & -1.07(09)\textsuperscript{\emph{c}} &  -1.12(18)\textsuperscript{\emph{c}}   \\
   &1080.26109 & 23 &  3 & 20 && 22 &  1 & 21 &&      						  &      					     & 							   &                			   &            						 &  -1.48(25)\textsuperscript{\emph{d}}   \\
   &1083.75583 & 27 &  3 & 24 && 26 &  1 & 25 &&      						  &      						 & 								  &                			   &            						 &  -0.38(08)\textsuperscript{\emph{d}}   \\
   &1092.31100 & 25 &  4 & 21 && 24 &  2 & 22 &&      						  &      						 & 							      &               			    &            						  &  -0.68(15)\textsuperscript{\emph{d}}   \\
   &1096.07712 & 37 &  4 & 33 && 36 &  2 & 34 &&      						  &      						 & 							      &                			   &            						 &  -1.66(31)\textsuperscript{\emph{d}}\B \\
\hline  \multicolumn{16}{l}{\T $^*$ This work; shifts induced by N$_2$ were not directly measured, but estimated using $\delta(\mathrm{N}_2)=(\delta(\mathrm{air})-0.21\cdot\delta(\mathrm{O}_2))/0.79$.}\\
        \multicolumn{16}{l}{$^a$ values according to HITRAN2016~\cite{GordonRothman:2017:The-HITRAN2016_a}, $^b$~Ref.~\cite{SokabeHammerich:1992:Photoacoustic_a}, $^c$~Ref.~\cite{SmithRinsland:1988:Measurements_b}, $^d$~Ref.~\cite{DeviBenner:1997:Air-Broadening_a}.}\\
\end{tabular}
\end{center}
\end{table*}

\begin{table*}[th!]
\caption{Comparison of measured and calculated room temperature pressure shift parameters $\delta$ for some lines in the $\nu_3$ band using He, Ar, Xe, O$_2$ and air as perturber gases. If not noted otherwise, measured values have been obtained in this work.\label{tab:mesureandcalc}}
\footnotesize
\begin{center}
\begin{tabular}{p{1mm}lp{2mm}p{2mm}p{2mm}p{1mm}p{2mm}p{2mm}p{2mm}p{0mm}llllllllllllll}
\hline
    \multicolumn{10}{l}{\T Transition$^a$}                                                                & \multicolumn{14}{l}{Pressure shift parameter $\delta$ (kHz / Pa)}       \\\cline{1-9}\cline{11-24}
\T    &Line position  &  \multicolumn{3}{l}{Upper state}   &&   \multicolumn{3}{l}{Lower state}                                && \multicolumn{2}{l}{He} && \multicolumn{2}{l}{Ar} && \multicolumn{2}{l}{Xe}& & \multicolumn{2}{l}{ O$_2$} &&  \multicolumn{2}{l}{air}\B \\\cline{3-5}\cline{7-9}\cline{11-12} \cline{14-15} \cline{17-18} \cline{20-21} \cline{23-24}
\T  &  \multicolumn{1}{l}{ (\wn) }  & $J^\prime$ & $K_a^\prime$ & $K_c^\prime$ && $J^{\prime\prime}$ & $K_a^{\prime\prime}$ & $K_c^{\prime\prime}$ && \multicolumn{1}{l}{meas} & \multicolumn{1}{l}{calc} && \multicolumn{1}{l}{meas} & \multicolumn{1}{l}{calc} && \multicolumn{1}{l}{meas} & \multicolumn{1}{l}{calc} && \multicolumn{1}{l}{meas} & \multicolumn{1}{l}{calc} && \multicolumn{1}{l}{meas} & \multicolumn{1}{l}{calc} \B \\ \hline
\T $A$	&1049.36312 &  9 &  1 &   8 &&  8 &  1 &  7 &&         & $-$0.36 && $-$0.54 & $-$0.73 && $-$0.85 & $-$0.92 && $-$0.54 & $-$0.61 && $-$0.50 & $-$0.64 \\
   $B$	&1049.62425 & 13 &  7 &   6 && 12 &  7 &  5 &&         & $-$0.36 && $-$0.64 & $-$0.73 && $-$0.83 & $-$0.91 && $-$0.59 & $-$0.62 && $-$0.69 & $-$0.69 \\
   $C$	&1052.04284 & 13 &  2 &  11 && 12 &  2 & 10 && $-$0.11 & $-$0.36 && $-$0.62 & $-$0.81 && $-$0.83 & $-$1.07 && $-$0.60 & $-$0.66 && $-$0.63 & $-$0.69 \\
   $D$	&1052.19324 & 24 & 11 &  14 && 23 & 11 & 13 && $-$0.13 & $-$0.36 && $-$0.72 & $-$0.92 && $-$0.98 & $-$1.43 && $-$0.58 & $-$0.77 && $-$0.57 & $-$0.84 \\
   		&1080.26109 & 23 &  3 &  20 && 22 &  1 & 21 &&         & $-$0.36 &&         & $-$0.97 &&         & $-$1.23 &&         & $-$0.81 && $-$1.48$^b$ & $-$0.82 \\
   		&1083.75583 & 27 &  3 &  24 && 26 &  1 & 25 &&         & $-$0.36 &&         & $-$0.98 &&         & $-$1.28 &&         & $-$0.83 && $-$0.38$^b$ & $-$0.83 \\
   	  	&1092.31100 & 25 &  4 &  21 && 24 &  2 & 22 &&         & $-$0.36 &&         & $-$0.98 &&         & $-$1.28 &&         & $-$0.83 && $-$0.68$^b$ & $-$0.84 \\
\B 		&1096.07712 & 37 &  4 &  33 && 36 &  2 & 34 &&         & $-$0.36 &&         & $-$1.05 &&         & $-$1.45 &&         & $-$0.86 && $-$1.66$^b$ & $-$0.86 \\ \hline
 \multicolumn{24}{l}{\T$^a$ values according to HITRAN2016~\cite{GordonRothman:2017:The-HITRAN2016_a}, $^b$ Ref.~\cite{SmithRinsland:1994:Measurements_a}.}\\
\end{tabular}
\end{center}
\end{table*}

\subsection{Line selection}\label{sec:selection}

The $\nu_3$ band of ozone at 10 $\upmu$m is the strongest fundamental and the ozone spectrum is particularly dense in the spectral region tuned by our QCL (see \figurename~\ref{fig:fig3}). The bottom panel of \figurename~\ref{fig:fig3} gives a zoom on the four lines for which we have measured the pressure shift. These lines have been chosen according to several criteria. First, lines were required to be free from interferences with other transitions. The interference between different transitions is a critical issue, especially at high pressures when pressure induced broadening leads to line overlapping and thus to changes of the observed line shapes, which impact the determination of the line center position.

Second, lines with different ``local quanta'' had been selected and the numbers are smaller than in the previous studies. This choice is important for probing a possible quantum number dependence of the pressure shift. Both criteria are well satisfied by the four lines marked \emph{A, B, C} and \emph{D} in Tables~\ref{tab:tab1} and \ref{tab:mesureandcalc}. For each of these lines, we list vacuum wavenumber, local quanta and measured pressure shift. As a comparison, previously measured line data \cite{SmithRinsland:1988:Measurements_b,SmithDevi:1997:Temperature_a,SokabeHammerich:1992:Photoacoustic_a} are also given in the same \tablename~\ref{tab:tab1}. Note that we have included line \emph{D} in our study despite its relative weak absorption, because it has been investigated in the past~\cite{SokabeHammerich:1992:Photoacoustic_a}.

\subsection{Uncertainty budget}\label{sec:uncertainties}
The shift coefficients are obtained from a linear fit of shift (or adjusted shifts) versus bath gas or total pressure. Uncertainties of pressure and wavenumber shifts were determined independently and the uncertainty has then been obtained from weighted fits with errors in both coordinates, using a total least squares algorithm that allows to introduce correlated uncertainty values \cite{AmiriSimkooei:2014:Estimation_a}. The implementation of the fit algorithm and its performance have been discussed elsewhere \cite{JanssenElandaloussi:2017:A-new-photometric_a}.
\subsubsection{Pressure}\label{sec:pressureunc}
Uncertainties in the pressure measurements have either been determined from manufacturer specifications that were verified using accepted procedures, or the pressure gauge has been calibrated over the relevant pressure range against traceable references.
The manufacturer specified accuracies for the high pressure sensor ($13 - 1000$\,hPa) with a nominal accuracy of 0.5\,\% take into account offset and gain drifts, non-linearity and hysteresis as well as the instrument resolution. The impact of ambient temperature variations, which have been kept within $\pm 1\,$K at the reference temperature of 20\,$^{\circ}$C, has also been considered according to the manufacturer specifications. Assuming a rectangular probability distribution of the manufacturer data, the corresponding standard measurement uncertainty has been derived from multiplication with the factor of $1/\sqrt{3}$:
\begin{equation}\label{eq:u_MKS122A}
u_{g}(p) = 15\,\mathrm{Pa} + 3.3\cdot 10^{-3} p.
\end{equation}
The gauge has been calibrated at 1000 hPa using a traceable high precision barometer (PTB210, \emph{Vaisala}, $u(p) \leq 5$\,Pa) and the gauge linearity has been verified through a series of adiabatic expansions. In this way, the compliance with manufacturer's specifications (Eq.~\ref{eq:u_MKS122A}) could be verified over the used pressure range.
The second gauge, which was used for quantification of ozone partial pressures, has a nominal accuracy of 0.08\,\%.
It has been calibrated in-house with respect to two similar gauges (MKS690 and MKS390 with 0.08 and 0.05\,\% nominal accuracies, respectively) which are both used as transfer standards that are regularly calibrated by the French national metrology institute LNE. The derived standard uncertainty of the stabilized gauge is
\begin{equation}\label{eq:u_MKS690}
u_{g_2}(p) = 0.016\,\mathrm{Pa} + 1.1\cdot 10^{-3} p,\quad   p \leq 13\, \mathrm{hPa}.
\end{equation}

In addition to the calibration related uncertainty, the finite resolution of the reading must be considered, as well as the fact that the PG partial pressure is obtained as a difference $p^{\prime} = p -   p^{m}_{\mathrm{O_3}}$ (see Eq.~({\ref{eq:p-shift3}})), where $p^{m}_{\mathrm{O_3}}$ has been fixed to the value of 47\,Pa with a rectangular uncertainty distribution of $\pm 20$\,Pa. Taking into account the reading related uncertainty (at a display resolution of 0.1 hPa), we obtain therefore
\begin{equation}\label{eq:p1}
u^2(p^{\prime}) = u_{g}^2(p^{\prime}+47\,\mathrm{Pa})) + (11.5\,\mathrm{Pa})^2 +  (10\,\mathrm{Pa})^2/12.
\end{equation}
The corresponding equation for the total pressure is
\begin{equation}\label{eq:p1}
u^2(p) = u_{g}^2(p) + (10\,\mathrm{Pa})^2/12 .
\end{equation}

\subsubsection{Frequency shift}\label{sec:shiftunc}
\begin{figure*}[t]
\centering
\includegraphics[width=14cm]{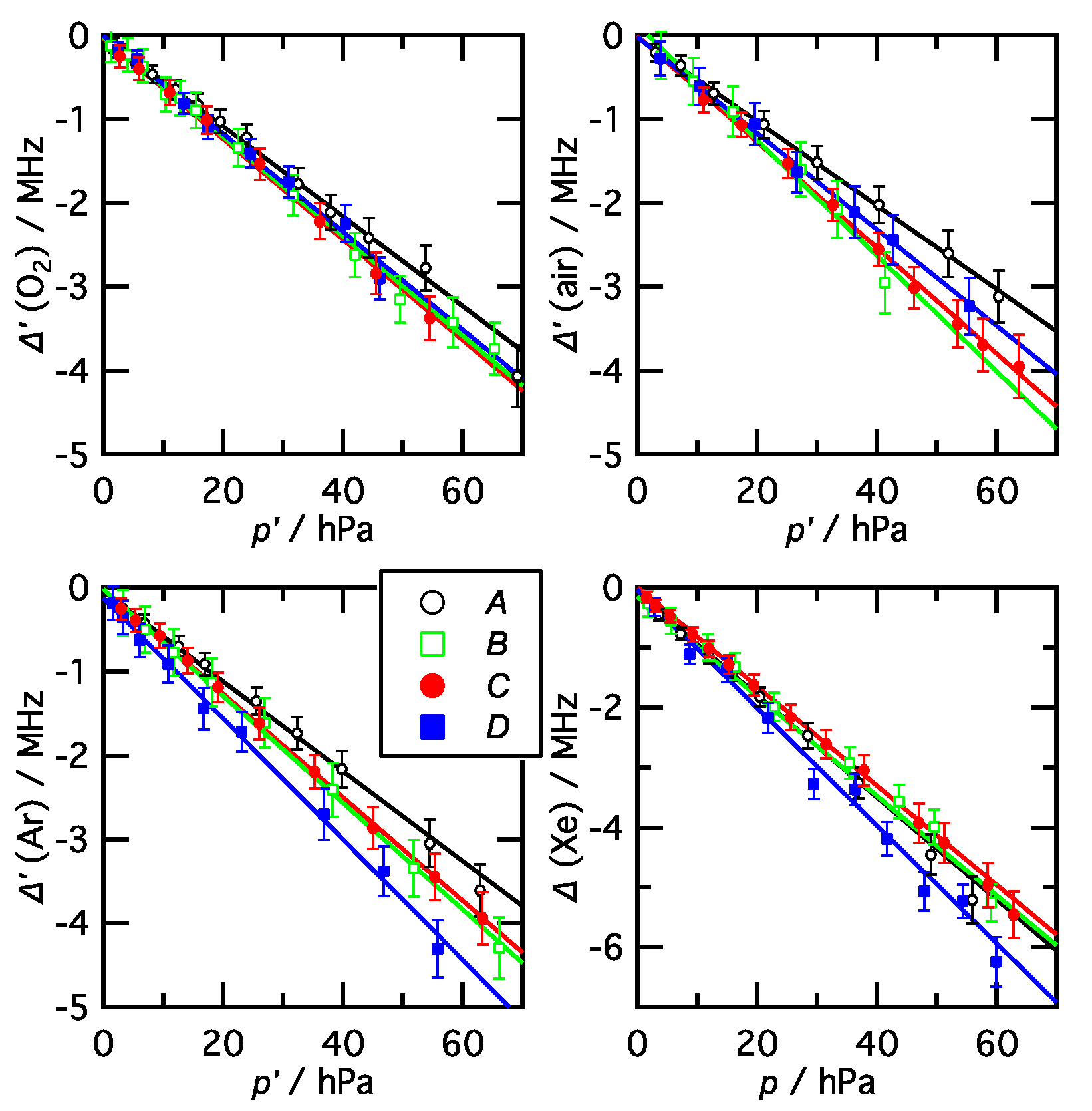}
\caption{Pressure shift of $A-D$ lines for four different perturber gases: O$_2$, air, Xe, and Ar. Error bars indicate standard uncertainties (at coverage a factor of $k=1$) in the measured quantities.\label{fig:fig4}}
\end{figure*}

Uncertainties in the frequency shift measurements derive from the calibration of the frequency scale, the non-linear laser frequency response to the tuning current and residual spectral etalon structures that impact the determination of the peak position and the reproducibility of the shift measurements. All of these factors are assumed to be independent. \\
The repeatability (rep) of an individual shift measurement has been determined from repeated spectral scanning for each value in Fig.~\ref{fig:fig4}. From these scans, a mean value, the standard deviation and the standard uncertainty of the mean have been derived.\\
The effect of residual structures (fr) which persist over the time scales of a shift measurement, however, cannot be estimated from the observed scatter. Its impact on the determination of the peak position has therefore been estimated independently by investigating selected spectra under varying experimental conditions. Similar to our peak finding algorithm, a quadratic function has been fitted to the centre region of the peaks such as shown in Fig.~\ref{fig:fig2a}. This yielded the typical peak shape at a given pressure as well as residual structures that were assumed to be characteristic for a given pressure, but would change position over the time. In order to take into account slow, but unknown changes of these structures, we have explored the variation of the fitted peak position as a function of the shift of the residual structure. By allowing continuous shifts over several fringe periods, the following pressure dependent uncertainty has been derived from the standard deviation of fitted peak  positions
\begin{equation}
u_{fr} = \sqrt{(0.06 + 0.0028 p/\mathrm{hPa})^2 + 0.06^2}\,\mathrm{MHz}.
\end{equation}
This expression takes into account that residual structures also impact the determination of the reference peak. In addition to the two above contributions, a 2\,\% relative uncertainty has to be considered. It takes into account the uncertainty of the frequency calibration and slight non-linearities of the frequency scale. This results in the standard uncertainty of a shift measurement given by
\begin{equation}
u^2 (\Delta) = u_{fr}^2  + u_{rep}^2 + (0.02 \Delta)^2
\end{equation}
The self-shift corrected shift (see Eq.~(\ref{eq:p-shift3})) also suffers from the contribution through the unknwown self shift coefficient $\delta_s$ and the difference in ozone partial pressures. The corresponding expression is
\begin{equation}\label{eq:uncshift2}
u^2 (\Delta^{\prime}) = u_{fr}^2  + u_{rep}^2 + (0.02 \Delta^{\prime})^2 + 267 \delta_s^2\,\mathrm{Pa}^2.
\end{equation}
For illustration purposes, we consider conservative values of $\delta_s$ in the range between $-0.5$ and $ -1.5$\,kHz/Pa, on the basis that self broading parameters are a few tens of percent larger than air broadening parameters. Even a value as large as $ -1.5$\,kHz/Pa implies an uncertainty related to the last term in Eq.~(\ref{eq:uncshift2}) of only 0.024\,MHz. This must be compared to other contributions to $u(\Delta^{\prime})$, such as $u_{fr}$, which varies between 0.089 and 0.26\,MHz over the relevant pressure range between 2 and 70\,hPa. Taking a more typical value of $\delta_s=-1.0\,$kHz/Pa, the self shift contribution is at least 4.5 times smaller than other terms and can thus almost be neglected in the quadratic addition.

\subsubsection{Slope modification factor}\label{sec:dilution}
Method 1 measurements on He and Xe (see Eq.~\ref{eq:p-shift_mixingratio}) retrieve a slope $a$ that slightly deviates from the shift parameter. The modification can be ascribed to the constant factor
\begin{equation}\label{eq:dilution}
	m = 1 -  \chi \left(1- \frac{\delta_s}{ \delta_g} \right),
\end{equation}
where $\chi $ denotes the ozone mixing ratio.
For the Xe measurements $\chi = 8  \cdot 10^{-3}$ has been determined from independent pressure measurements. Assuming  $1\leq \delta_s/\delta_g\leq 2 $, we obtain
\begin{equation}\label{eq:uncm}
	m = 1.004 \pm 0.003
\end{equation}
wich needs to be taken into account when determining the shift coefficient from $\delta_g = a / m$.
\subsection{Correlations}\label{sec:correlations}
Correlations impact on uncertainties derived from the linear fit of shift versus pressure. Here we give correlaion coefficients between $x$-values, $y$-values and different $x$-$y$ pairs, from which the covariance matrix for the weighting of the linear fit can be set up. For a more detailed discussion, we refer to Ref.~\cite{JanssenElandaloussi:2017:A-new-photometric_a}.
\subsubsection{First method: He and Xe}
Pressure readings $p$ are likely correlated through calibration errors~\cite{JanssenElandaloussi:2017:A-new-photometric_a}. We conservatively assume
\begin{equation}\label{eq:corrp1}
	r(p_i,p_j) = \frac{u_g(p_i)u_g(p_j)}{u(p_i)u(p_j)}
\end{equation}
for runs at different pressures $p_i$ and $p_j$. Individual shift measurements are clearly correlated by the scaling error. Residual structures may also lead to correlation between different measurements of the line shift. Without further detailed information, we assume a correlation coefficient of 0.5 for this particular contribution. This leads to the overall expression
\begin{equation}\label{eq:corrshift1}
r(\Delta_i,\Delta_j) = \frac{4 \cdot 10^{-4} \Delta_i \Delta_j + 0.5\cdot {u_{fr}(\Delta_i)}u_{fr}(\Delta_j)}{u(\Delta_i)u(\Delta_j)}
\end{equation}
for different measurements $i, j$ of the shift.

\subsubsection{Second method: Ar, O$_2$, and air}
As before, we conservatively assume
\begin{equation}\label{eq:corrp2}
	r(p_i^\prime,p_j^\prime) = \frac{u_g(p_i)u_g(p_j)}{u(p^\prime_i)u(p^\prime_j)}
\end{equation}
where the total pressure  $p_i$ is offset from the PG pressure $p_i^\prime$ by the ozone partial pressure $p^m_\mathrm{O_3}=47\,$Pa in the measurement cell. For the correlation between corrected shifts and the correlation between pressure and shifts we obtain
\begin{eqnarray}\label{eq:corrshift2}
r(\Delta_i^\prime,\Delta_j^\prime) &=& \frac{4 \cdot 10^{-4} \Delta_i^\prime \Delta_j^\prime + 0.5\cdot {u_{fr}(\Delta_i^\prime)}u_{fr}(\Delta_j^\prime) + \delta_s^2\, u^2(p^r_{\mathrm{O_3}})}{u(\Delta_i^\prime)u(\Delta_j^\prime)}  \\
	r(\Delta_i^\prime,p_i^\prime) &=& \delta_s\frac{133\,\mathrm{Pa}^2}{u(\Delta_i^\prime)u(p_i^\prime)}.
\end{eqnarray}
For calculation of the last two terms, the self pressure broadening parameter has been approximated by $\delta_s = -1.0$\,kHz/Pa.
\section{Discussion}\label{sec:discussion}
\subsection{Bathgas dependence}\label{sec:bathgas}
\figurename~\ref{fig:fig3} shows the experimental results obtained for the four lines marked in \tablename~\ref{tab:tab1}. Data are obtained using five different perturber gases: O$_2$, air, He, Ar, and Xe. For He, only pressure shifts of lines $C$ and $D$ are presented in \tablename~\ref{tab:tab1}, because shifts of the other two lines gave values comparable to the error bars. We have thus also omitted to show a graph of $\Delta$(He). The pressures of the perturber gases were chosen depending on the experimental conditions, but generally spanned the 2 to 70\,hPa pressure range.

As can be seen from \tablename~\ref{tab:tab1}, the four lines studied in the present work all show a negative shift as a function of pressure and the size of the shift strongly depends on the nature of the perturber. For example, line \emph{C} presents shift parameters of $-0.13$\,kHz/Pa for He, $-0.72$\,kHz/Pa for Ar, and  $-0.98$\,kHz/Pa for Xe. This bath gas dependence of $\left|\delta(\mathrm{He})\right| <\left|\delta(\mathrm{Ar})\right| < \left|\delta(\mathrm{Xe})\right|$ has been observed for all lines, as listed in \tablename~\ref{tab:tab1}. Such a behaviour has already been observed using molecules other than ozone, such as HCl~\cite{Be61}, CO~\cite{Be94}, H$_2$~\cite{He96}, and HBr or HI~\cite{Do07} and is confirmed by our semiclassical calculations using the lowest order expression of the complex RB formalism~\cite{LynchGamache:1996:Fully_a}.


\subsection{Air-shifts}\label{sec:airshifts}
\begin{figure}[t]
\centering
\includegraphics[width=0.9\columnwidth]{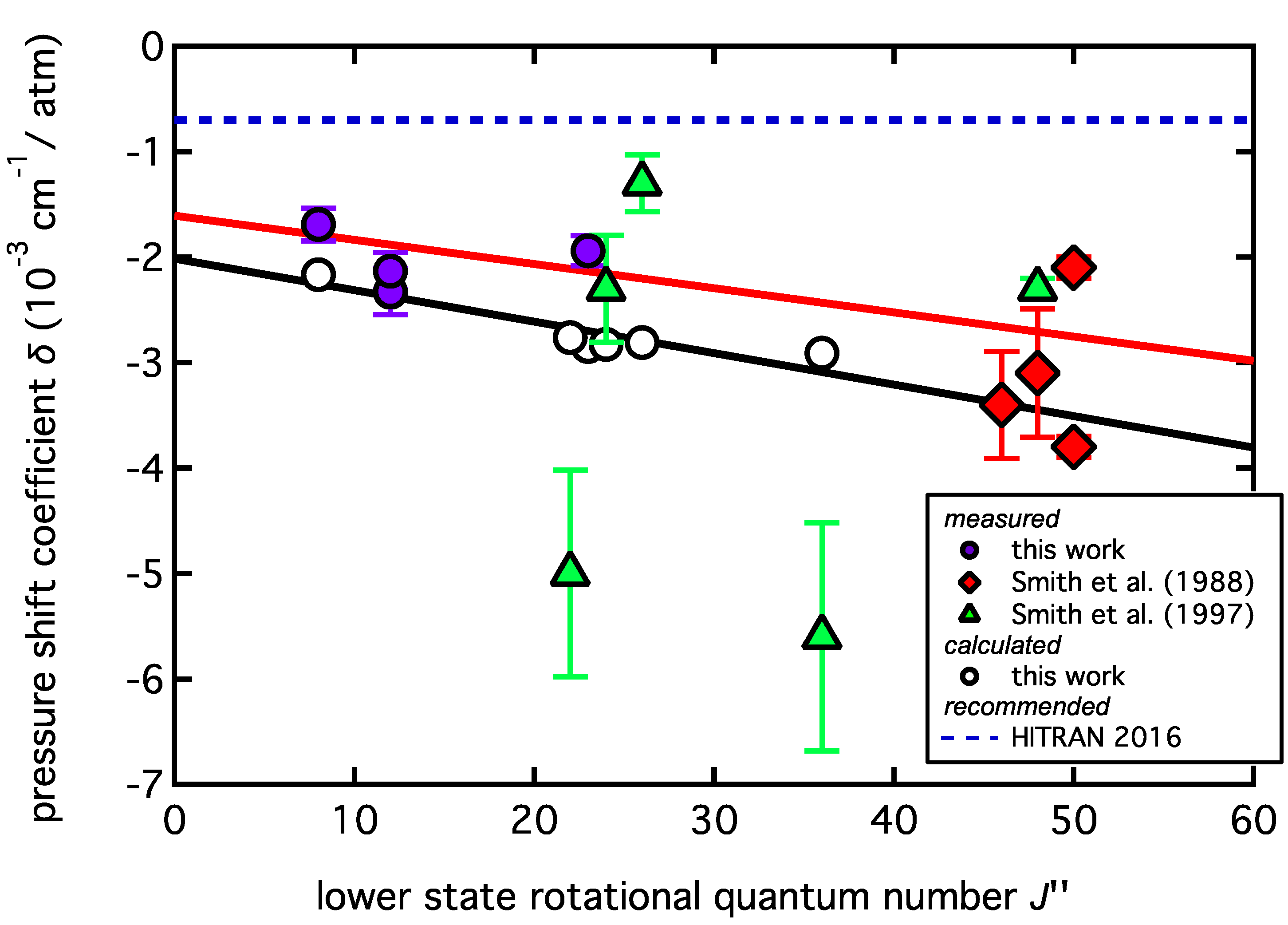}
\caption{ Calculated and measured pressure shift coefficients $\delta$ versus $J^{\prime\prime}$ in the $\nu_3$ band of ozone (open and closed symbols, respectively). Measured data are taken respectively from Ref.~\cite{SmithRinsland:1988:Measurements_b} (red diamonds), Ref.~\cite{SmithDevi:1997:Temperature_a} (green triangles), and this work (violet circles). The dashed line on top indicates  the current recommendation (HITRAN2016, \cite{GordonRothman:2017:The-HITRAN2016_a}). Black open circles are calulated shift coefficients in this work. The red and black solid lines respectively indicate a weighted linear fit to the measurements and a simple linear fit to the calculated data.} \label{fig:fig5}
\end{figure}

Our measurements of $\delta$(air) of lines $A - D$ range from $-0.50$ to $-0.69$\,kHz/Pa, while values of $\delta$(O$_2$) are between $-0.54$ and $-0.60$\,kHz/Pa, as can be inferred from \tablename~\ref{tab:tab1}.
In order to compare with other studies, we have estimated the pressure shift induced by N$_2$ molecules by assuming a simple additive law between O$_2$ and N$_2$ shifts:
\begin{equation}\label{eq:N2shift}
\delta(\mathrm{N}_2)=\frac{\delta(\mathrm{air})-0.21\delta(\mathrm{O}_2)}{0.79}.
\end{equation}
It must be noted that based on their measurements at 3\,$\upmu$m Smith {et al.}~\cite{SmithRinsland:1994:Measurements_a} caution against this assumption. Given the scatter of these data, however, it is somewhat unclear whether this observation is really significant, especially because pressure broadening parameters observed in the same study actually comply with the additivity rule.
We notice that our results are comparable with the FTS measurements of Smith~\emph{et al.}~\cite{SmithRinsland:1988:Measurements_b,SmithDevi:1997:Temperature_a}, but do not seem to agree with the photoacoustic laser study of Sokabe \emph{et al.}~\cite{SokabeHammerich:1992:Photoacoustic_a}: Line $D$ presents a pressure shift four times the size of our result and the sole \emph{P}-branch transition at 1037.43368\,\wn shows a positive shift. Despite not having measured in the $P$-branch, a direct comparison with these results is also complicated by the fact that the authors did not report uncertainty values for their measurements. Therefore, and due to the fact that all other measurements using air or N$_2$ not only have lower values than Ref.~\cite{SokabeHammerich:1992:Photoacoustic_a}, but are also much closer to our results, especially if assigned uncertainties are in the 0.1 kHz/Pa range or lower (see  \tablename~\ref{tab:tab1}  and \figurename~\ref{fig:fig5}), we have excluded the results of Sokabe~\emph{et al.} in the following discussion. Note that without identifying neither the vibrational band nor the individual transitions, Sokabe~\emph{et al.} \cite{SokabeHammerich:1992:Photoacoustic_a} measure shifts between 0 and $-0.9$\,kHz/Pa (or 0 and $-0.003$\,\wn /atm) for the vast majority of lines reported in Ref.~\cite{SmithRinsland:1988:Measurements_b}. This seems to be largely compatible with our result and the measurements in Refs.~\cite{SmithRinsland:1988:Measurements_b} and \cite{SmithDevi:1997:Temperature_a}.

\begin{figure}[t]
\centering
\includegraphics[width=\columnwidth]{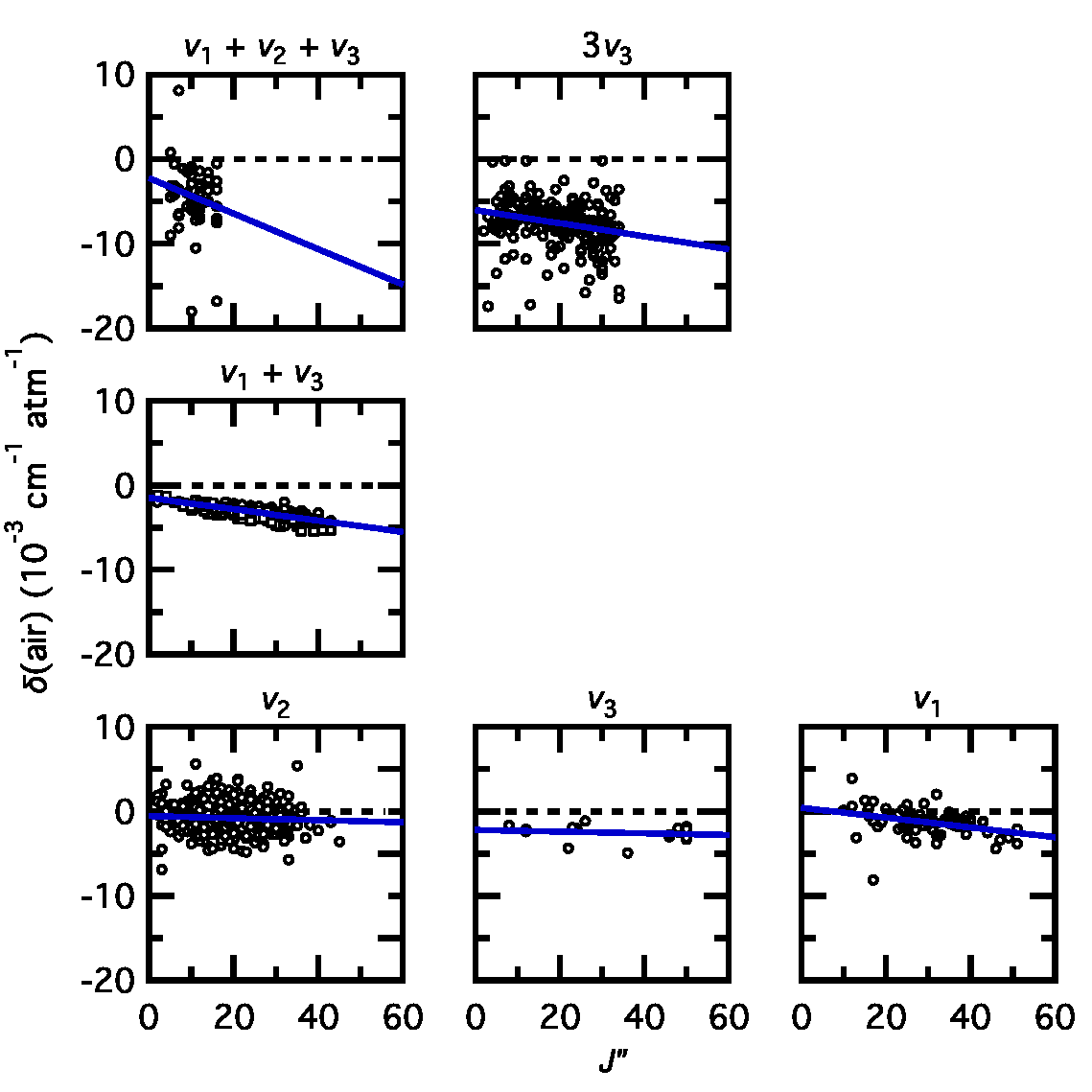}
\caption{Comparison of air-shift measurements as a function of $J^{\prime\prime}$ in different ozone vibrational bands. Panels are arranged in order of increasing energy (left to right) and number of vibrational quanta (bottom to top). $\nu_1$: \cite{SmithRinsland:1988:Measurements_b}; $\nu_2$: \cite{DeviBenner:1997:Air-Broadening_a}; $\nu_3$: this work, \cite{SmithRinsland:1988:Measurements_b}, \cite{SmithDevi:1997:Temperature_a}; $\nu_1 + \nu_3$:  N$_2$ (circles) and O$_2$ (squares) \cite{BarbeBouazza:1991:Pressure_a}, ; $\nu_1 +\nu_2+ \nu_3$: \cite{SmithRinsland:1994:Measurements_a}; $3 \nu_3$: \cite{SmithRinsland:1994:Measurements_a}. Fit lines were obtained by direct fit to the data, except for the $\nu_1$ and the $\nu_1 + \nu_3$ bands. For $\nu_1 + \nu_3$, N$_2$ and O$_2$ were fitted separately and then combined using respective abundance weights of 0.79 and 0.21.} \label{fig:fig6}
\end{figure}

As already laid out in Section~\ref{sec:selection}, one of the criteria for selecting lines was the ability to span a significant range of rotational quantum numbers. Quantum numbers of $J^{\prime\prime}$ between 8 and 50 and $K_c^{\prime\prime}$ between 7 and 47 are now comprised when we include previous FTS measurements \cite{SmithRinsland:1988:Measurements_b,SmithDevi:1997:Temperature_a} in the analysis. \figurename~\ref{fig:fig5} shows the air shift coefficients listed in \tablename~\ref{tab:tab1} as functions of $J^{\prime\prime}$.
We observe a general increase in absolute values of pressure shift coefficients as rotational quantum numbers increase. Solid lines shown in \figurename~\ref{fig:fig5} represent linear fits to experimental and calculated results. With the exception of the two transitions  $(23, 3, 20) \leftarrow (22, 1, 21)$ at 1080.26098\,\wn and $(37, 4, 33) \leftarrow (36, 2, 34)$ at 1096.07705\,\wn, that suffer from relatively high measurement uncertainties when compared to the remainder of the data, the general linear trend is quite well followed. Note that the Pearson correlation coefficient $r$ for the linear fit is $\sim -0.2$ when all lines are admitted to the fit, and $\sim -0.5$ when these two lines are excluded. Results of these fits can be used to estimate the rotational quantum number dependence of the pressure shift. We obtain
\begin{eqnarray}\label{eq:airshiftJ}
\delta / (\mathrm{kHz / Pa}) &=&-0.45(12)-0.008(3)\cdot  J^{\prime\prime} \quad \mathrm{or} \nonumber \\
\delta / (10^{-3}\mathrm{cm^{-1} / atm}) &=&-1.5(4)-0.027(11)\cdot  J^{\prime\prime}
\end{eqnarray}
when a unweighted fit is made through all data except the two above points.
Numbers in parentheses give standard deviations in units of the last digit obtained from the present least-squares analysis. A very similar result is obtained from a weighted fit to all data points. The fact that shifts are negative and that the effect increases in magnitude with increasing quantum number has already been observed in other bands (Fig.~\ref{fig:fig6}) and this is now confirmed for the $R$-branch of the $\nu_3$ fundamental~\cite{SmithRinsland:1988:Measurements_b,SmithDevi:1997:Temperature_a,DeviBenner:1997:Air-Broadening_a,BarbeBouazza:1991:Pressure_a,SokabeHammerich:1992:Photoacoustic_a,SmithRinsland:1994:Measurements_a,GamacheArie:1998:Pressure-broadening_a,LavrentievaOsipova:2009:Calculations_a}. Fig.~\ref{fig:fig6} also implies that the magnitude of the shift increases with vibrational excitation.

Fig.~\ref{fig:fig5} gives shift coefficients in \wn/atm, from which it might be inferred that the average shift in the $R$-branch of the $\nu_3$ band is $-2.8\cdot 10^{-3}$\,\wn/atm ($-0.84\,$ kHz/Pa) when all data are included (see Table~\ref{tab:shift-per-band}). One obtains  $-2.5\cdot 10^{-3}$\,\wn/atm if the $(23, 3, 20) \leftarrow (22, 1, 21)$ and $(37, 4, 33) \leftarrow (36, 2, 34)$ transitions are neglected or if a weighted average of all data points is taken. The derived numbers can thus be seen as representative. While our analysis certainly suffers from a small number of data points, it already suggests that typical shift coefficients for atmospheric applications are between $-2$ and $-3\cdot 10^{-3}$\,\wn/atm, and thus are about four times larger than the current recommendation in the HITRAN2016 database, which proposes a value of $-0.7\cdot 10^{-3}$\,\wn/atm for all transitions belonging to the ozone $\nu_3$-band. A unique value of $\delta = -3\cdot 10^{-3}$\,\wn/atm, or better the linear relation in Eq.~(\ref{eq:airshiftJ}), is much more representative, at least as far as $R$-branch transitions are concerned. The data also imply a slight and global $J$ dependence, as found for other bands (see Table~\ref{tab:shift-per-band} or Fig.~\ref{fig:fig6}). It would thus be even more appropriate to take include this dependence into the databases.

\section{Summary and Outlook}\label{sec:conclusions}
We have measured the pressure shifts of ozone molecular lines using a free-running distributed-feedback QCL emitting at 9.54\,$\upmu$m. The pressure shift coefficients of four intense rovibrational transitions in the $R$-branch of the $\nu_3$ fundamental band of ozone, induced by O$_2$, air and the noble gases He, Ar, and Xe could be obtained. Negative pressure-shift coefficients have been found with values ranging between $-0.11$ and $-0.13$\,kHz/Pa for He, between $-0.54$ and $-0.72$\,kHz/Pa for Ar, and between $-0.82$ and $-0.98$\,kHz/Pa for Xe. This finding corresponds to the usual correlation with the polarizability of the perturber gas, as confirmed by a simple calculation using the semi-classical formalism of Robert and Bonamy.

Air and oxygen resulted in  pressure shift coefficients between $-0.50$ and $-0.69$\,kHz/Pa. These values are close to previous results  \cite{SmithRinsland:1988:Measurements_b,SmithDevi:1997:Temperature_a} for other transitions in the same branch of the $\nu_3$ band. With the four new low $J$ and $K$ transitions, the available data now span the range between 8 and 50 in $J^{\prime\prime}$ and from 7 to 47 in $K_c^{\prime\prime}$, allowing to derive an empirical correlation between the shift coefficient and these quantum numbers. While suffering from a small number of data points, the analysis already suggests that shift coefficients for atmospheric applications are close to $-3\cdot 10^{-3}$\,\wn/atm, and thus are about four times larger than the current recommendation at $-0.7\cdot 10^{-3}$\,\wn/atm. A unique value of $\delta = -\cdot 10^{-3}$\,\wn/atm, or even a simple linear relation $\delta / (10^{-3}\mathrm{cm^{-1} / atm}) =-1.5(4)-0.027(11)\cdot  J^{\prime\prime}$, are much more representative for transitions in the $\nu_3$ band, especially for the $R$-branch which has been reviewed here. A negative $J^{\prime\prime}$ dependence of similar magnitude is also found using the simple semiclassical RB approach.

A systematic study using the present technique would be very useful to make up for the lack of shift parameters for $\nu_3$ lines of ozone. This seems to be an urgent matter as not even a single line in the $P$-branch has been measured as of today and as only one line of the $Q$-branch has a shift coefficient associated so far. An obvious extension of the current work is the study of the temperature dependence of shift parameters or the inclusion of different isotopomers, such as in~\cite{Su98}.

A subsequent stabilization of the QCL onto an optical frequency comb, which is intended for the very near future, will open up the possibility to perform metrological measurements of Doppler-free molecular lines and thus greatly enhance the sensitivity and accuracy of the measurements. As a consequence, much weaker lines will become accessible and the pressure shifting in the intense $\nu_3$-band can be studied more systematically.

\begin{table*}
	\caption{Comparison of air induced line shift parameters in different vibrational bands.}\label{tab:shift-per-band}
	\scriptsize
	\begin{tabular}{lr@{ -- }lld{1}d{1}d{1}d{0}rl}\hline
	\T\B Band  &  \multicolumn{2}{l}{Span in $J^{\prime\prime}$} & Data- &   \multicolumn{1}{l}{HITRAN}  & \multicolumn{1}{l}{Average}  &  \multicolumn{3}{l}{Linear Fit\ $^{a}$ } &  Reference with spectral \\ \cline{7-9}
	\T & \multicolumn{2}{l}{} &points& \multicolumn{1}{l}{($10^{-3}$\,cm$^{-1}$\,atm$^{-1}$)} & \multicolumn{1}{l}{($10^{-3}$\,cm$^{-1}$\,atm$^{-1}$)} & \multicolumn{1}{l}{offset} & \multicolumn{1}{l}{slope } & \multicolumn{1}{l}{Pearson $r$ } &  resolution (cm$^{-1}$)\ $^{b, c}$\\
	\B & \multicolumn{2}{l}{} &&& & \multicolumn{1}{l}{($10^{-3}$\,cm$^{-1}$\,atm$^{-1}$)} & \multicolumn{1}{l}{($10^{-6}$\,cm$^{-1}$\,atm$^{-1}$)} &  & \\
	 \hline	
	\T$\nu_2$ 			&  2 & 45	& 362	& -0.8	& -0.78	& -0.54	& -13	& $-0.06$ 	& \citealp{DeviBenner:1997:Air-Broadening_a}: 0.005\\
	$\nu_3$ 				&  8 & 50 &  13	& -0.7	& -2.8	& -2.2 	& -10 	& $-0.17$  	& TW: 0.0002, \citealp{SmithRinsland:1988:Measurements_b}: 0.005, \citealp{SmithDevi:1997:Temperature_a}: 0.005\\
	$\nu_1$				& 10 & 56	&  64	& -0.7	& -1.3	& 0.4 	& -58 	& $-0.34$ 	& \citealp{SmithRinsland:1988:Measurements_b}:  0.005, \citealp{SmithDevi:1997:Temperature_a}: 0.005 \\
	$\nu_1 + \nu_3\ {}^d$	&  2 & 43	& 138	& -3 	& -2.9	& -1.5 	& -67 	& $-0.85$ 	& \citealp{BarbeBouazza:1991:Pressure_a}: 0.002\\
	$\nu_1 + \nu_2 + \nu_3$	&  5 & 16	&  54	&  0 	& -4.5	& -2.3	& -210 	& $-0.18$  	& \citealp{SmithRinsland:1994:Measurements_a}: 0.01\\
	\B $3\nu_3$    		&  2 & 34 & 202	& -8 	& -7.5	& -6.0 	& -77	& $-0.23$  	& \citealp{SmithRinsland:1994:Measurements_a}: 0.01\\\hline
	\multicolumn{10}{l}{\T$^a$ Due to the low number of available datapoints, stated uncertainties have been used as weights for fitting shifts in the $\nu_3$ band.}\\
	\multicolumn{10}{l}{$^b$ TW -- this work}\\
	\multicolumn{10}{l}{$^c$ Spectral resolution in parantheses given as FWHM-width of laser line (TW) or the resolution of FTIR instrument.}\\
	\multicolumn{10}{l}{$^d$ Values obtained for O$_2$ and N$_2$ and then combined using abundance weights: $x(\mathrm{air}) = 0.79 x(\mathrm{N}_2) + 0.21 x(\mathrm{O}_2)$.}
	\end{tabular}
	
\end{table*}
%

%
%
%

\section*{Acknowledgement}
This work was supported by grants from Région Ile-de-France in the framework of the DIM ACAV and by the LABEX Cluster of Excellence FIRST-TF (ANR-10-LABX-48-01), within the Program "Investissements d'Avenir" operated by the French National Research Agency (ANR). We thank F.~Thibout, P.~Marie-Jeanne, C.~Rouill\'{e}, and H.~Elandaloussi for their generous and skillfull help in setting up the experiment. We would like to express our gratitude to Y.~T\'{e} for very fruitful discussions and we deeply acknowledge L.~Hilico for providing a low-noise current source for driving the QCL.

\end{document}